\documentclass[11pt]{article}
\usepackage[margin=1in]{geometry}
\usepackage{amsmath,amssymb}
\usepackage{booktabs,array}
\usepackage{enumitem}
\usepackage{microtype}
\usepackage{graphicx}
\graphicspath{{figures/}}
\usepackage[backend=biber,style=numeric-comp,sorting=none,maxbibnames=6]{biblatex}
\usepackage[hidelinks]{hyperref}
\usepackage{caption}
\usepackage{setspace}
\newcommand{\EBV}{Epstein--Barr virus}
\newcommand{\MS}{multiple sclerosis}

\title{\textbf{Stage-Resolved EBV-Induced B-Cell Programs Across CNS Selection and Multiple Sclerosis Compartments}}
\author{Hasan M. Jamil\\
\small Department of Computer Science, University of Idaho, USA\\
\normalsize Alexander Gow\\
\small Center for Molecular Medicine and Genetics, Wayne State University, USA}
\date{}

\begin{document}
\maketitle

\noindent
\textbf{Contact Author:} Hasan Jamil, jamil@uidaho.edu

\begin{abstract}
\textbf{Background and Objectives:}
Epstein--Barr virus (EBV) is strongly implicated in multiple sclerosis (MS), but it is unknown whether EBV-conditioned B-cell states remain detectable as stable RNA programs after CNS selection. We tested transcriptional persistence, the specificity of cross-system molecular bridges, and whether independently defined EBV programs show stage-dependent downstream behavior.

\textbf{Methods:}
We sequentially reanalyzed public datasets spanning experimental EBV infection (GSE189141), EBV-associated CNS-homing B cells in humanized mice (GSE299939), prospective early-MS blood (GSE267750), paired human CSF/blood (GSE133028), and an independent paired cohort (GSE138266). A frozen EBV-induced factor (F12) was transferred without outcome-driven modification. Post hoc controls tested lineage and exposure specificity, direct brain anatomy, sequencing depth/dropout, coherence-aware nulls, metabolic confounding, external EBV programs, and a disjoint-gene sensitivity.

\textbf{Results:}
F12 did not predict later MS activity (AUC=0.487, exact $p=0.489$) and was lower in CSF in all 9 paired participants (mean $-0.0560$), surviving coherence-matched calibration (two-sided empirical $p=0.015$) and hypoxia/glycolysis residualization. In 5 EBV mice, F12 remained lower in brain after matched-depth rarefaction (5/5 negative, $p=0.03125$) and relative to a mouse coherence-matched null ($p=0.0225$). The raw TBX21$^+$CXCR3$^+$ fraction attenuated after depth matching, so the splenic gate was not treated as evidence of anatomical depletion. A 287-gene homing-to-CSF bridge was strongly enriched but showed no B-lineage advantage and was substantially reproduced in PBS mice. In human CSF, a variance-qualified early-EBV program was depleted (mean $-0.0492$, $p=0.0197$), whereas Carter latency-III and LCL programs increased ($+0.0666$, $p=0.0092$; $+0.0148$, $p=0.0096$). Removing all genes shared by the early and LCL source signatures preserved opposite directions (early-only $-0.0815$, $p=0.0237$; LCL-only $+0.0295$, $p=0.0013$). No external program met the frozen prospective criterion.

\textbf{Discussion:}
The data support stage- and compartment-dependent transcriptional remodeling rather than a stable EBV RNA barcode. A marker-defined trafficking phenotype need not preserve the upstream transcriptomic state, and statistically strong cross-system resemblance need not establish molecular ancestry. EBV-directed biomarker and intervention studies should combine host RNA with virologic, clonotype, antigen-specific, epigenetic, and longitudinal measurements.
\end{abstract}

\section{Introduction}

The association between \EBV{} and \MS{} now rests on convergent epidemiologic and mechanistic evidence. In a large longitudinal military cohort, EBV seroconversion preceded MS and was associated with a marked increase in subsequent disease risk, while neurofilament light increased only after EBV infection \cite{Bjornevik2022}. EBV-directed B-cell responses can cross-react with GlialCAM, supporting molecular mimicry \cite{Lanz2022}, and EBV infection can reshape antigen presentation on the MS-risk HLA-DR15 background \cite{Wang2026}. EBV-reactive clonally expanded T cells have also been identified in MS CSF \cite{Hayashi2026}.

B cells are a plausible cellular carrier of an EBV-related process. Early EBV infection generates heterogeneous host B-cell trajectories \cite{SoRelle2022}, and host transcription evolves across early proliferation and lymphoblastoid transformation \cite{Price2012}. Carter et al. likewise identified latency-III-associated host changes in transformed B-cell models \cite{Carter2002}, raising the possibility that ``EBV conditioning'' is not a single transcriptomic state but a sequence of states. In a humanized-mouse model, EBV expands oligoclonal T-bet$^+$CXCR3$^+$ B cells that enter the CNS and attract inflammatory T cells; B-cell depletion or CXCR3 blockade reduces CNS infiltration \cite{Laderach2025}. A related TBX21$^+$CD21$^{lo}$CXCR3$^+$ phenotype is expanded in early human MS, although direct EBV infection of those human cells was not demonstrated \cite{SoRelle2025}.

These observations motivate a mechanistic question with direct biomarker consequences: if EBV conditions a B-cell population relevant to CNS entry, should an EBV-induced host transcriptional program remain detectable in later blood or CSF? A positive answer would make static RNA signatures attractive bridges between infection and disease. A negative answer would not refute EBV causality, because lineage, antigen specificity, epigenetic memory, and altered response potential can persist while transcription is remodeled by proliferation, endothelial transit, tissue entry, cytokine exposure, and antigen encounter. Contemporary EBV--MS models therefore accommodate persistent-driver and transient or ``hit-and-run'' mechanisms \cite{Wahbeh2025,Mizzi2026}.

\begin{figure}[t]
\centering
\includegraphics[width=\textwidth]{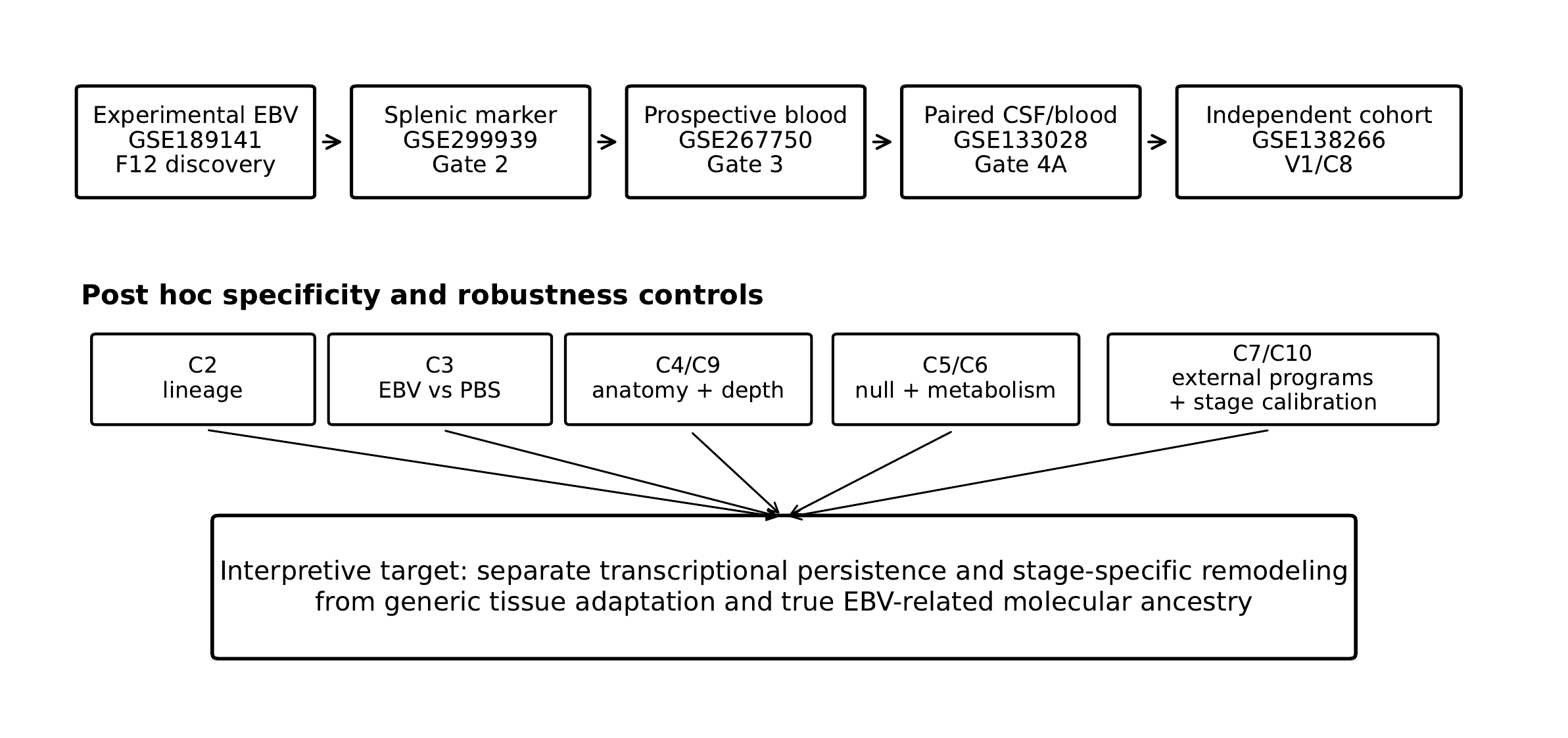}
\caption{Sequential analysis and specificity/robustness controls. Frozen discovery and transfer analyses preceded the post hoc controls. C2--C10 and a final disjoint-gene sensitivity test lineage, exposure, anatomy, sequencing depth, module coherence, metabolic confounding, external EBV programs, and independent-cohort evaluability. The controls constrain interpretation without redefining the original gates.}
\label{fig:architecture}
\end{figure}

The inferential danger is that a lost initiating program can be replaced post hoc by a broader cross-system overlap and then interpreted as evidence of molecular ancestry. We therefore designed the analysis to permit falsification. We first tested a frozen experimentally EBV-induced B-cell factor across an EBV-linked homing phenotype, prospective human disease, and paired human CSF/blood. When static persistence failed, we characterized a broader gene-level bridge, then challenged the bridge and F12 with anatomy, lineage, exposure, sequencing-depth, coexpression, metabolic, and external-program controls (Figure~\ref{fig:architecture}). The final analysis distinguishes 3 possibilities that are often conflated: persistence of the same RNA state, stage-specific transcriptional remodeling, and generic tissue adaptation.

\section{Methods}

\subsection{Datasets, sequence, and inferential units}
We reanalyzed five public single-cell studies in a fixed sequence (Table~\ref{tab:datasets}). The biological unit was donor for experimental infection, mouse for experimental CNS analyses, and participant for human comparisons; cells were never treated as independent inferential replicates. Candidate states and gate rules were frozen before transfer. Post hoc specificity controls were introduced only after the original persistence and bridge analyses were complete.

\begin{table}[t]
\centering
\caption{Datasets and roles.}
\label{tab:datasets}
\small
\begin{tabular}{p{1.8cm}p{3.2cm}p{3.2cm}p{6.6cm}}
\toprule
Dataset & System & Biological unit & Role \\
\midrule
GSE189141 & Primary human B cells after experimental EBV infection & Donor ($n=2$) & Discover EBV-induced host factors; held-out-cell consistency within the same donor libraries \cite{SoRelle2022}. \\
GSE299939 & Human B cells from humanized mice & Mouse (3 marker-defined; 5 EBV and 3 PBS direct-anatomy evaluable) & Gate 2 marker phenotype; direct brain-versus-spleen analysis; anatomical surrogate and depth/dropout tests \cite{Laderach2025}. \\
GSE267750 & Longitudinal early-MS/CIS PBMCs & Participant ($n=16$) & Prospective baseline and longitudinal F12 tests; external EBV-program tests \cite{SoRelle2025}. \\
GSE133028 & Paired human CSF/blood & Participant ($n=9$ evaluable MS/CIS) & Primary compartment test; cross-system bridge; lineage, coherence, metabolic, and stage-calibrated external-program controls \cite{Ramesh2020}. \\
GSE138266 & Independent paired CSF/PBMC cohort & Participant (5 MS, 5 controls; 4 MS primary evaluable) & Untouched module challenge; relaxed $\ge10$-B-cell disease-specificity sensitivity recovered 1 control \cite{Schafflick2020}. \\
\bottomrule
\end{tabular}
\end{table}

\subsection{Stage 1, held-out cells, and Gate 2}
In GSE189141, a deterministic balanced sample of QC-passing cells from 2 donors at days 0, 2, 5, and 8 was decomposed with a fixed 12-factor MiniBatchNMF. Candidate ranking used cross-donor temporal concordance, overlap with prespecified nuisance programs (type-I/II interferon, TNF/NF-$\kappa$B, cell cycle, stress/DNA damage, plasmablast), and nuisance-residualized late effects. F12, F3, and F5 were frozen. F12's top genes included GBE1, BCL2, HIF1A-AS2, P4HA1, HILPDA, PFKFB4, BNIP3L, SLC2A3, BACH2, CD22, and FCRL3. Cells excluded from discovery but from the same 2 donor libraries were then scored unchanged; this was a held-out-cell consistency check, not independent replication.

Gate 2 transferred only the 3 frozen factors to splenic human B cells from GSE299939. Target cells were TBX21$>0$ and CXCR3$>0$; each mouse required at least 25 target and 250 other B cells. A factor passed if all 3 mice had positive target-minus-other mean difference and AUC$>0.5$.

\subsection{Human prospective and paired-compartment tests}
In GSE267750, outcome-blind high-confidence B-cell gating was used in 13 participants with later MS activity/diagnosis and 3 long-term-no-activity participants. The protected test was baseline F12 difference with rank AUC and exact label permutation. In GSE133028, the canonical B-cell gate required at least 2 markers among CD19, MS4A1, CD79A, CD79B, CD37, and CD22 and a B-cell score exceeding T-cell, myeloid, and NK scores. Participants required at least 20 B cells in both CSF and blood. The protected effect was participant-level mean F12$_{CSF}-$F12$_{blood}$ with exact sign-flip inference.

\subsection{Cross-system bridge and module challenge}
For the original transition analysis, stage effects were donor-level EBV late-minus-day0 expression ($E_1$), marker-defined splenic TBX21$^+$CXCR3$^+$ minus other B cells ($E_2$), and participant-level CSF-minus-blood B-cell expression ($E_3$). Common genes were independently rank-normalized within stage. Core UP required $Z\ge1.0$ and unit-level directional consistency. The conceptual bridge was $E_2$-UP $\cap$ $E_3$-UP with $E_1$ unconstrained. The 287-gene core bridge was frozen before pathway/network characterization. Its strongest actin/adhesion/trafficking community (Module 1) was tested unchanged in GSE138266 against immune Module 6 and generic translation and oxidative-phosphorylation Modules 4/5.

\subsection{Post hoc specificity and robustness controls}
C2 recomputed human CSF-minus-blood effects in B cells, CD4 T cells, and monocytes on a common 5,326-gene universe while holding experimental $E_2$ fixed. C3 replaced the marker-defined $E_2$ with direct brain-minus-spleen B-cell effects from the source study's deposited Seurat object (Zenodo DOI 10.5281/zenodo.15602185), requiring at least 20 cells per organ, and repeated the bridge in EBV and PBS mice \cite{Laderach2025}.

C4 tested, within the same direct-anatomy object, whether brain selection enriched the TBX21$^+$CXCR3$^+$ fraction while preserving F12. C5 generated expression/detection- and coherence-matched modules for the human F12 contrast. C6 computed Hallmark HYPOXIA and GLYCOLYSIS scores excluding all F12 genes, residualized F12 against those scores, and repeated the human and mouse compartment tests.

C7 froze 3 outcome-independent external EBV programs before downstream scoring: a 27-gene modern-symbol subset of the 38 latency-III-up RNAs reported by Carter et al. from a 4,146-feature array \cite{Carter2002}, and early-EBV UP40 and LCL UP40 programs derived from the independent GSE20200 resting-B/early-proliferating/LCL series \cite{Price2012}. Hallmark TNF/NF-$\kappa$B was a mechanistic comparator. C8 repeated the GSE138266 module analysis with a labeled secondary evaluability threshold of 10 rather than 20 B cells per compartment.

C9 addressed sequencing-depth/dropout in the 5 frozen evaluable EBV mouse pairs. We reported median UMI counts and detected genes by organ, greedily matched brain and spleen cells within mouse on standardized log UMI and detected-gene counts, and independently thinned each matched pair to the lower UMI depth. Two hundred fixed-seed rarefactions were performed without selecting a favorable replicate. We repeated the binary TBX21$^+$CXCR3$^+$ fraction, a continuous TBX21/CXCR3 expression score, and F12. A separate sensitivity restricted cells above the pooled 25th percentile of detected genes. Finally, an expression/detection- and coherence-matched mouse null tested whether F12 depletion exceeded generic brain-to-spleen module remodeling. Thirty-seven of 40 F12 genes mapped to the mouse object; the main cross-platform score renormalized frozen weights over mapped genes, with the original 40-gene denominator retained as a sensitivity.

C10 calibrated the C7 human CSF directions symmetrically. For each external EBV program, random modules were matched on expression and detection and retained only if their within-compartment residual coherence was within the frozen tolerance of the observed program. The stage-pattern rule was frozen before viewing C10: the pattern was called calibrated only if the early program was significantly negative and both Carter and LCL programs significantly positive by two-sided empirical $p\le0.05$. After C10, a reviewer-requested disjoint-gene sensitivity applied the unchanged C10 matching, coherence tolerance, candidate count, and empirical-test procedure to the early-only and LCL-only genes after removing every gene shared by the 2 frozen source signatures. This sensitivity was explicitly post hoc and no threshold was changed after its result was viewed.

\subsection{Primary research question and Classification of Evidence}
The primary prospective research question was whether baseline F12 distinguished participants with subsequent MS activity/diagnosis from LTNA in GSE267750. The source cohort was retrospective and narrow-spectrum, so the evidence is not rated above Class III. However, F12 was learned and frozen in an external experimental dataset, the B-cell gate was defined without outcome information, no outcome-dependent fitting or feature selection was performed in GSE267750, and the protected baseline comparison and exact label-permutation inference were specified before group testing. On that basis, this study provides Class III evidence that baseline F12 did not discriminate subsequent MS activity/diagnosis from LTNA in the analyzed public early-MS/CIS cohort (AUC=0.487, exact $p=0.489$).

\subsection{Standard Protocol Approvals, Registrations, and Patient Consents}
This study is a secondary analysis of deidentified, publicly available datasets and involved no new recruitment, intervention, or access to identifiable private information. Ethical approvals and informed-consent procedures were those of the original studies and are reported in the source publications. No additional participant consent was obtained for this public-data reanalysis.

\section{Results}

\subsection{F12 was experimentally induced but did not persist into prospective blood or human CSF}
F12 was the strongest discovery candidate by the original specificity ranking: cross-donor temporal $\rho=1.0$, no top-40 overlap with the frozen nuisance panels, and positive nuisance-residualized late effects in both donors. However, its held-out-cell late effects fell to 0.123 and 0.068 with temporal $\rho=0.60$, making F12 the smallest held-out effect among F12/F3/F5. Thus it was transferable but already weaker outside the discovery subset.

In Gate 2, F12 was higher in TBX21$^+$CXCR3$^+$ splenic B cells in all 3 mice (AUCs 0.854, 0.736, 0.625); F5 also passed 3/3, so the gate was not uniquely selective for F12. In prospective human blood, baseline F12 was essentially unrelated to later MS activity (mean MS-LTNA difference $-0.000363$, AUC=0.487, exact $p=0.489$), about $0.016$ participant SD in magnitude. In paired human samples F12 was lower in CSF than blood in all 9 participants (mean $-0.0560$, median $-0.0554$), the opposite of the protected persistence hypothesis.

\begin{figure}[t]
\centering
\includegraphics[width=\textwidth]{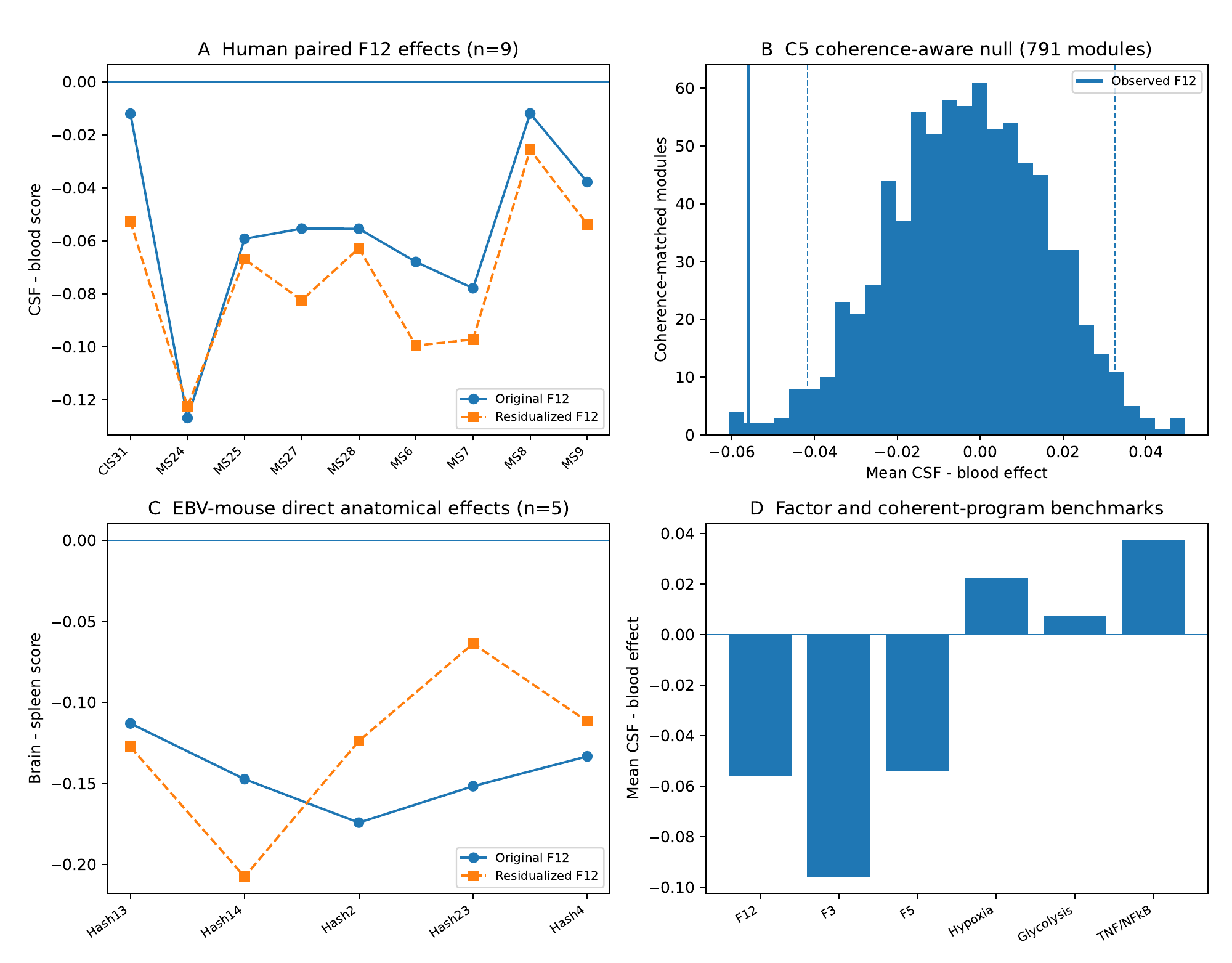}
\caption{F12 loss is robust to coherence and metabolic controls. (A) Participant-level human CSF-minus-blood F12 effects before and after hypoxia/glycolysis residualization. (B) C5 coherence-matched null; the vertical solid line is observed F12 and dashed lines are the empirical 2.5th/97.5th percentiles. (C) Direct EBV-mouse brain-minus-spleen F12 effects before and after residualization. (D) Mean human compartment effects for the discovered factors and coherent Hallmark benchmarks.}
\label{fig:f12}
\end{figure}

The CSF decrease survived the stronger C5 calibration. F12 coherence was 0.189; 791 synthetic modules matched within the frozen $\pm0.025$ band. Their mean effect was $-0.00335$ (95\% empirical interval $[-0.0417,0.0325]$), whereas F12 remained $-0.0560$ (lower-tail $p=0.00758$, two-sided $p=0.0152$; Figure~\ref{fig:f12}). F3 and F5 were also negative, whereas Hallmark hypoxia, glycolysis, and TNF/NF-$\kappa$B were not, arguing against a generic property of coherent modules.

F12 contained conspicuous metabolic/hypoxia-associated genes that had not been explicit discovery nuisances. C6 therefore residualized F12 against nonoverlapping Hallmark HYPOXIA and GLYCOLYSIS scores. The human decrease persisted in 9/9 participants and became $-0.0737$ (one-sided exact $p=0.00195$). The corresponding Hallmark compartment effects were slightly positive rather than negative. Thus F12 loss was not reducible to broad hypoxia/glycolysis variation.

\subsection{The original gene bridge was statistically strong but failed module specificity}
The original core bridge contained 287 $E_2$-UP/$E_3$-UP genes versus 142.8 expected (OR=3.11, $p\approx1.2\times10^{-39}$); 120 were also EBV-up in $E_1$. Excluding RPL/RPS genes strengthened rather than erased the bridge (278 observed vs 131.4 expected; OR=3.36; $p\approx4.6\times10^{-43}$), so ribosomal content did not account for the overlap. The bridge contained a coherent 34-gene actin/adhesion/trafficking module enriched for cytoskeletal and leukocyte-migration biology.

\begin{figure}[t]
\centering
\includegraphics[width=\textwidth]{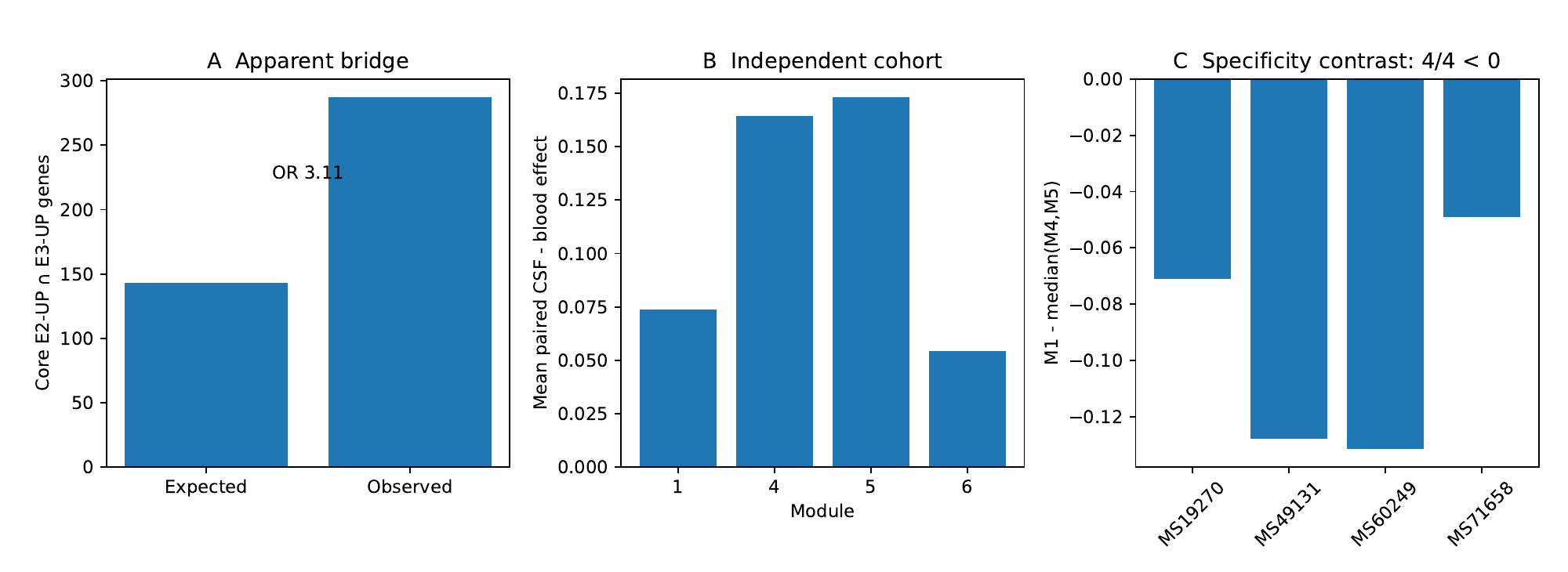}
\caption{Original bridge and independent module challenge. The 287-gene marker-defined homing-to-CSF bridge was strongly enriched and contained coherent trafficking biology, but its prespecified Module-1 specificity contrast was negative relative to generic translation/OXPHOS modules in every evaluable independent MS pair.}
\label{fig:bridge}
\end{figure}

Yet in the independent GSE138266 challenge, only 4 MS pairs met the frozen $\ge20$-B-cell rule, making the original strong-pass $p\le0.05$ criterion mathematically unattainable. Module 1 was positive in 3/4 (mean 0.0737), but translation Module 4 was positive in 4/4 (mean 0.164) and OXPHOS Module 5 in 3/4 (mean 0.173). Module 1 minus the median generic comparator was negative in 4/4 (mean $-0.0949$; Figure~\ref{fig:bridge}). C8 lowering the evaluability threshold to 10 recovered only 1 of 5 controls; that control's Module-1 effect (0.0833) slightly exceeded the MS mean, leaving disease specificity non-evaluable rather than supported.

\subsection{Lineage controls showed no B-cell advantage}
\begin{figure}[t]
\centering
\includegraphics[width=0.86\textwidth]{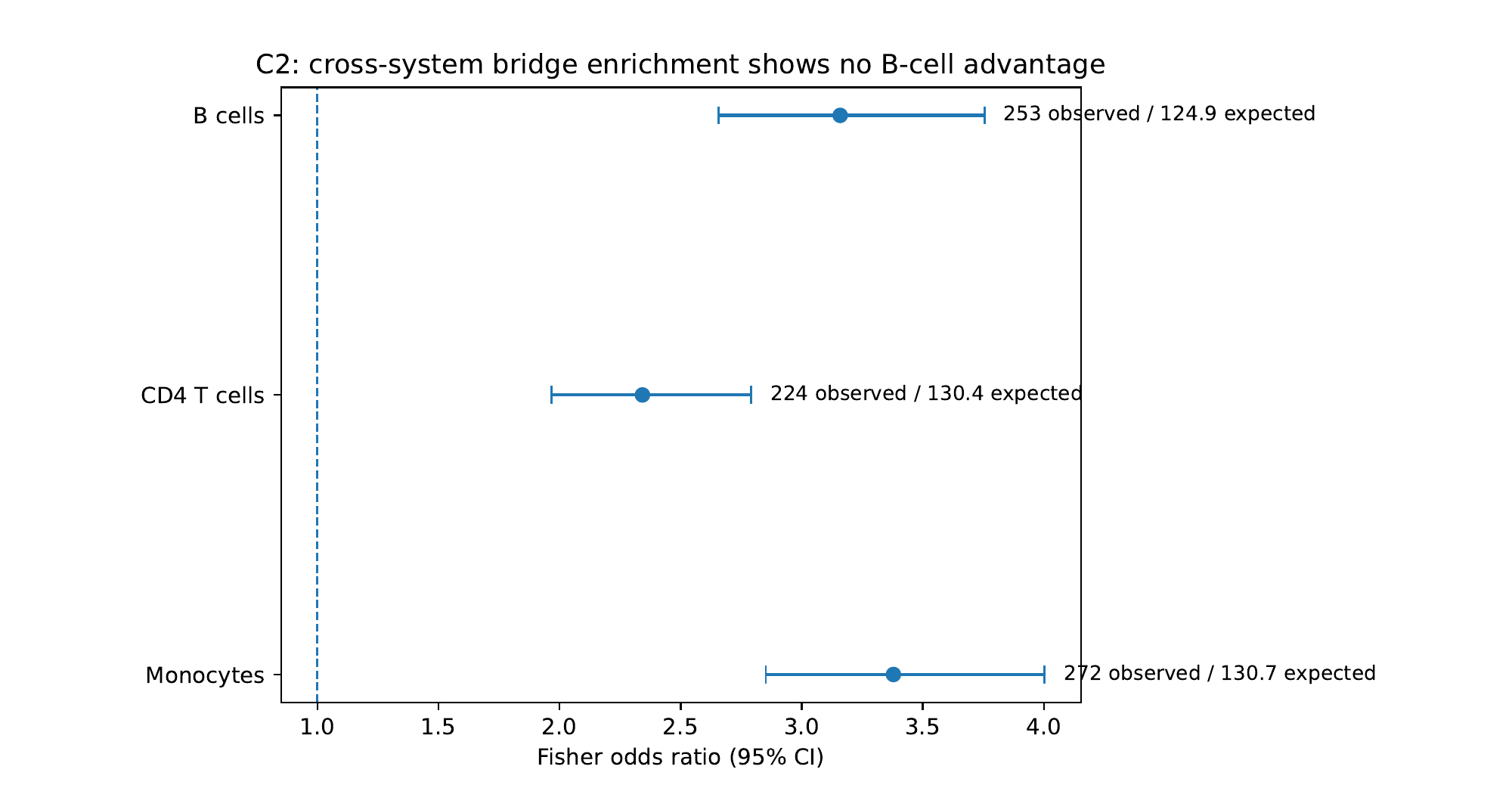}
\caption{C2 lineage specificity control. Fisher odds ratios and 95\% Wald confidence intervals are shown on the identical common-gene universe. All lineages exhibit strong enrichment; the analysis supports no B-cell-specific advantage.}
\label{fig:lineage}
\end{figure}

On the common 5,326-gene universe, the same experimental B-cell $E_2$ contrast aligned strongly with human CSF-minus-blood effects in B cells (253 observed vs 124.9 expected; OR 3.16, 95\% CI 2.66--3.75), CD4 T cells (224 vs 130.4; OR 2.34, 95\% CI 1.97--2.79), and monocytes (272 vs 130.7; OR 3.38, 95\% CI 2.85--4.00). The intervals overlap and there was no B-cell advantage (Figure~\ref{fig:lineage}). The bridge therefore captures a substantial compartmental program shared across lineages rather than a B-cell-specific downstream continuation.

\subsection{Direct anatomy separated trafficking evidence from transcriptomic-state persistence}
Replacing the marker-defined contrast with actual brain-minus-spleen B-cell effects preserved a statistical bridge to human CSF: 233 EBV genes versus 121.5 expected (OR=2.73, $p=9.18\times10^{-28}$). But PBS mice also produced a strong bridge (239 vs 141.8; OR=2.24, $p=7.94\times10^{-20}$). On the 6,080-gene common universe, EBV and PBS brain-minus-spleen effects correlated at $\rho=0.767$; 174 of 232 EBV bridge genes were shared with PBS (75\%, Jaccard=0.586). Thus direct anatomical selection generated a real cross-system correspondence that was not EBV-specific.

\begin{figure}[t]
\centering
\includegraphics[width=\textwidth]{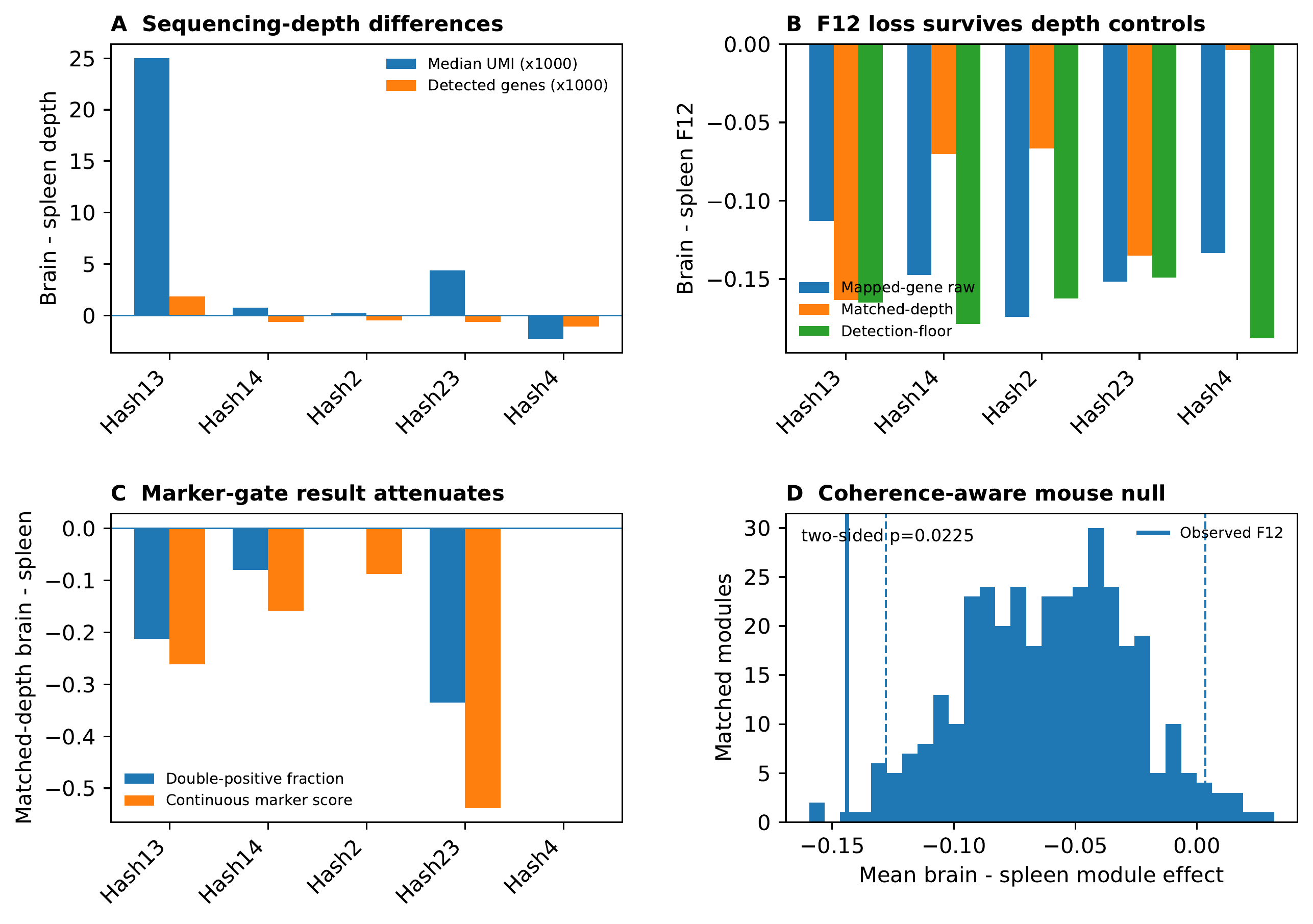}
\caption{C9 direct-anatomy depth robustness. (A) Brain-minus-spleen differences in median UMI depth and detected genes for the 5 frozen evaluable EBV mice. (B) F12 remains lower under mapped-gene raw scoring, matched-depth rarefaction, and the detection-floor sensitivity. (C) The binary TBX21$^+$CXCR3$^+$ result attenuates after matched-depth processing; the continuous marker score is shown in parallel. (D) F12 is more negative than the coherence-aware mouse null (two-sided empirical $p=0.0225$).}
\label{fig:direct}
\end{figure}

F12 itself was lower in brain in all 5 evaluable EBV mice. C9 showed that this result was not explained by shallower brain sequencing. Median UMI depth was not lower overall in brain (mean paired median difference $+5,621$ UMIs; one-sided test for lower brain depth $p=0.875$), although detected-gene complexity was lower in 4/5 mice (mean paired difference $-192$ genes, $p=0.406$). After within-mouse matching on UMI and detected-gene complexity and pairwise rarefaction to equal UMI depth, F12 remained lower in the median rarefaction for all 5 mice (mean of mouse medians $-0.0877$, one-sided $p=0.03125$). A gene-detection-floor sensitivity was again negative in all 5 (mean $-0.1685$, $p=0.03125$). The mouse coherence-aware null was shifted negative, consistent with generic tissue remodeling, but observed mapped-gene F12 ($-0.1438$) remained more depleted than matched modules (null mean $-0.0613$, 95\% interval $[-0.1280,0.0035]$, two-sided empirical $p=0.0225$; Figure~\ref{fig:direct}).

The marker phenotype was less robust. The raw TBX21$^+$CXCR3$^+$ fraction was lower in 4/5 mice, but after matched-depth rarefaction the mean binary difference was $-0.1254$ with one-sided $p=0.125$. A continuous TBX21/CXCR3 score remained lower in 4/5 matched-depth mouse summaries (mean $-0.2086$, $p=0.0625$) and was lower in all 5 after the detection-floor sensitivity ($p=0.03125$). We therefore do not infer anatomical depletion of the double-positive population from scRNA-seq. Instead, the result shows that the splenic double-positive gate is not a reliable transcriptomic surrogate for the state recovered from brain.

This distinction is important because the source study's flow cytometry, clonotype, imaging, and perturbational evidence supports CNS trafficking of T-bet$^+$CXCR3$^+$ B cells \cite{Laderach2025}. Our reanalysis does not challenge that trafficking conclusion; it tests the narrower claim that a splenic RNA gate preserves the upstream molecular state after anatomical selection. It does not.

Thirty-seven of 40 F12 genes mapped to the mouse object. Renormalizing frozen weights over mapped genes yielded the reported mean $-0.1438$; retaining the original 40-gene denominator with the 3 unmapped genes contributing zero yielded $-0.1295$. All 5 EBV mice were negative under either convention.

\subsection{Human CSF showed calibrated stage-structured EBV-program remodeling}
C7 asked whether F12 generalized beyond the 2-donor NMF discovery. In the original C7 scoring, 29/40 early-UP40 and 29/40 LCL-UP40 genes were available downstream. The 27-gene Carter latency-III subset increased in human CSF (mean $+0.0666$), the early-EBV program decreased (mean $-0.0424$), and the LCL program increased (mean $+0.0142$). In EBV mouse brain-versus-spleen, Carter was nearly null ($-0.0010$), early-EBV increased ($+0.0398$), and LCL increased ($+0.0905$), so the ordering was not identical across species.

C10 additionally required nonzero variance in the within-compartment pseudobulk profiles used to define coherence. This variance-qualified universe retained all 27 Carter genes, 25/40 early genes, and 28/40 LCL genes; that restriction explains the small C7-to-C10 effect changes. Carter remained positive (observed $+0.0666$ vs null 95\% interval $[-0.0483,0.0498]$, two-sided empirical $p=0.0092$), early-EBV remained negative ($-0.0492$ vs $[-0.0400,0.0256]$, $p=0.0197$), and LCL remained positive ($+0.0148$ vs $[-0.0026,0.0114]$, $p=0.0096$). The LCL increase was therefore calibrated but modest in absolute magnitude compared with F12 loss ($|0.0148|$ vs $|0.0560|$); the Carter increase was similar in magnitude to F12 loss.

\begin{figure}[t]
\centering
\includegraphics[width=\textwidth]{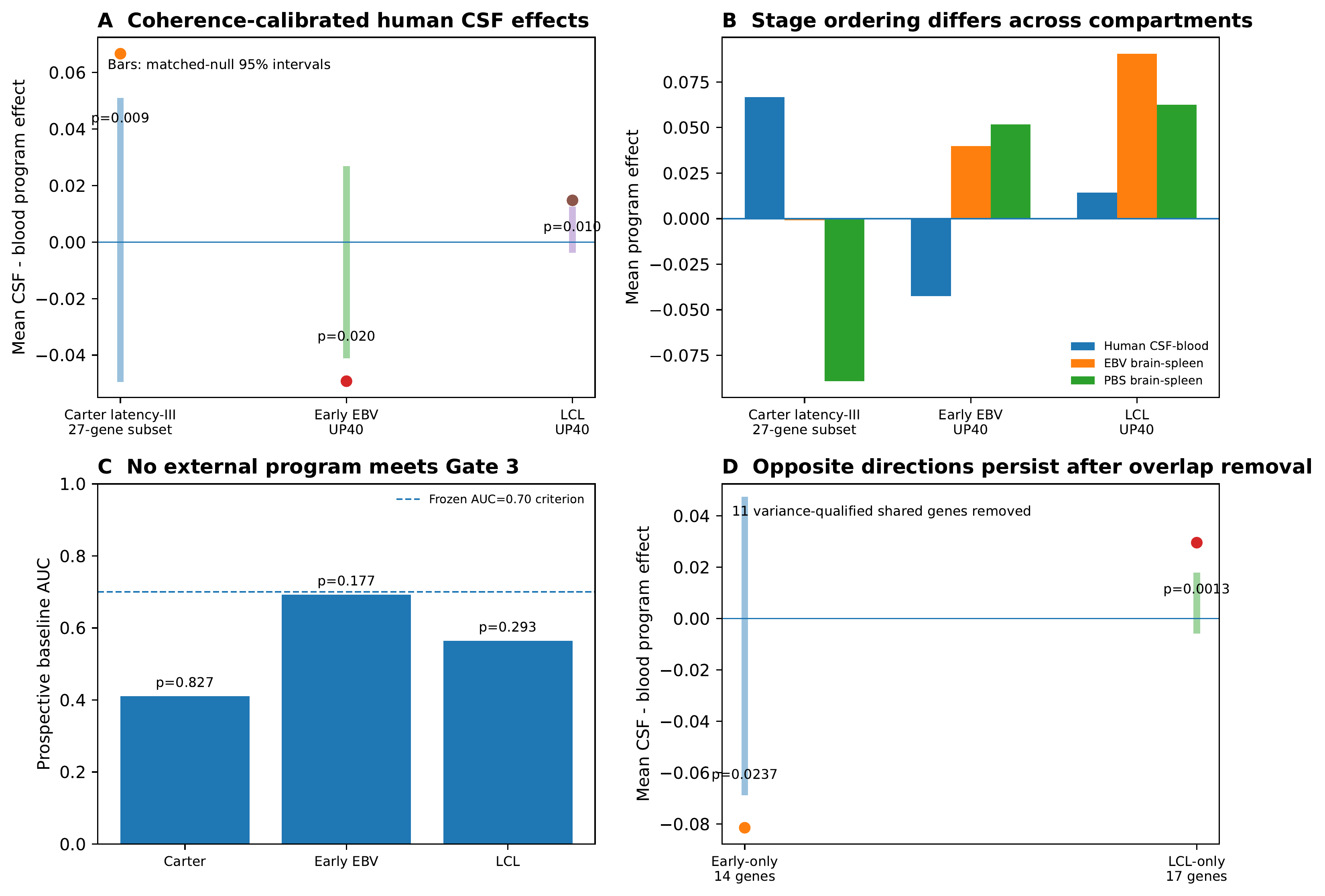}
\caption{C7/C10 external EBV-program analysis. (A) Human CSF-minus-blood effects relative to each program's coherence-matched null. The LCL increase is statistically calibrated but smaller in absolute magnitude than F12 loss. (B) Raw C7 effects across human CSF and mouse brain/spleen show that stage ordering is not identical across species. (C) No external EBV program meets the frozen prospective AUC=0.70 criterion in GSE267750. (D) Disjoint-gene sensitivity after removing all early/LCL source overlap: 14 variance-qualified early-only genes remain negative and 17 LCL-only genes remain positive relative to their own unchanged coherence-matched nulls.}
\label{fig:external}
\end{figure}

The 2 GSE20200 source signatures shared 22/40 genes. Within the C10 variance-qualified universe, 11 of those shared genes remained, leaving 14 evaluable early-only genes and 17 evaluable LCL-only genes. Applying the unchanged C10 null procedure to these disjoint sets preserved the opposite directions: early-only mean CSF-minus-blood was $-0.0815$ (null 95\% interval $[-0.0676,0.0461]$, two-sided empirical $p=0.0237$), whereas LCL-only was $+0.0295$ (null 95\% interval $[-0.00463,0.01658]$, $p=0.00125$). Thus the human stage ordering was not generated by the genes shared between the early and LCL source signatures (Figure~\ref{fig:external}).

The frozen C10 criterion for a calibrated human stage pattern was met. In human CSF, an early/proliferative EBV program is depleted while later transformation/latency-III-like programs are retained or increased. However, the mouse external-signature directions prevent a universal stage law of CNS selection: F12 loss is independently robust in mouse anatomy, whereas the external early-UP40 program rises there. None of the external programs met the frozen prospective Gate-3 criterion; the closest remained early-EBV UP40 (AUC=0.692, exact $p=0.177$).

\section{Discussion}

\subsection{Acute-program loss and stage-structured remodeling are distinct biological findings}
The central result is not that EBV lacks a role in MS. The epidemiologic, antigenic, HLA, trafficking, and clonotype evidence for EBV operates independently of F12 \cite{Bjornevik2022,Lanz2022,Laderach2025,Wang2026,Hayashi2026}. Rather, the analysis shows that a particular experimentally EBV-induced host B-cell program is not a durable molecular label of the downstream population. F12 was induced experimentally and associated with a splenic phenotype linked to CNS homing, yet it was essentially unrelated to prospective MS activity, fell in every paired human CSF sample, and fell again in every evaluable EBV-mouse brain. Human and mouse coherence-aware nulls, metabolic residualization, matched-depth rarefaction, and a detection-floor sensitivity all preserved that conclusion.

C10 adds a second biological result. The loss does not apply to EBV-conditioned transcription indiscriminately. In human CSF, an independently derived early/proliferative program was depleted, whereas transformation-associated LCL and latency-III programs were increased relative to their own coherence-matched nulls. The LCL effect was small ($+0.0148$) relative to F12 loss ($-0.0560$), so it should not be read as equally large biological displacement. More importantly, removing every gene shared by the early and LCL source signatures strengthened rather than erased the separation: early-only genes remained negative ($-0.0815$) and LCL-only genes positive ($+0.0295$), each outside its own matched null. The distinction is biologically plausible because EBV-driven B-cell outgrowth is itself a temporal state transition rather than a single response \cite{Price2012,Carter2002}. The result therefore favors a model of \emph{transcriptional remodeling}: an acute EBV-induced state can be erased while later EBV-conditioned or transformation-associated programs remain detectable.

That inference must remain compartment- and model-specific. The external early-UP40 program increased rather than decreased in EBV mouse brain, while Carter was nearly null. Accordingly, C10 establishes stage structure in human CSF, not a universal cross-species rule. What is conserved across human CSF and mouse brain is the loss of F12 specifically, not the direction of every early EBV signature.

The durable biological record of an upstream event may therefore reside in a different molecular channel from the one that first revealed it: BCR clonotype, antigen specificity, chromatin state, differentiation history, survival circuitry, or altered response thresholds. This creates a direct experimental prediction. Longitudinal lineage-resolved multi-omic studies should observe discontinuity of the acute RNA state alongside continuity of at least one ancestry-bearing or functional feature if true state handoff occurs.

\subsection{A trafficking phenotype is not a transcriptomic surrogate}
The depth-control analysis changes how the anatomical result should be stated. We no longer interpret the raw 4/5 decrease in the TBX21$^+$CXCR3$^+$ fraction as evidence that this population is depleted from brain. That binary statistic is dropout-sensitive and attenuated after matched-depth rarefaction. The source study's flow cytometry, BCR clonotype, imaging, depletion, and CXCR3-blockade experiments provide a more direct basis for the trafficking conclusion \cite{Laderach2025}.

Our finding is orthogonal: the splenic TBX21$^+$CXCR3$^+$ gate does not preserve the F12 transcriptomic state in anatomically recovered brain B cells. Trafficking competence, marker phenotype, anatomical selection, and transcriptomic state are therefore distinct observables. This is a generalizable lesson for single-cell mechanism studies: a marker-defined population can be causally relevant to migration while still undergoing major state remodeling after tissue entry.

\subsection{Cross-system resemblance is not molecular ancestry}
The original 287-gene bridge was highly nonrandom and biologically coherent. Yet its mechanistic interpretation failed two specificity tests that are rarely applied to cross-system transcriptomic bridges. The bridge had no B-lineage advantage, and direct brain-to-CSF correspondence was substantially reproduced in PBS mice. These are not failures of statistical power: they are positive demonstrations that a statistically strong bridge can arise from compartment biology shared across lineages and exposures.

A significant overlap can reject independence but cannot identify why two states resemble one another. Migration, cytoskeletal remodeling, tissue stress, nutrient availability, and immune activation can create convergence without shared causal ancestry. Exposure-negative animals and lineage-negative cells are therefore mechanistic controls, not optional embellishments. This principle generalizes beyond EBV--MS whenever transcriptomic similarity is used to connect an experimental perturbation to a downstream human disease compartment.

\subsection{Medical significance: implications for biomarkers and intervention studies}
The medical implication is that the timing and biological phase of the measurement matter. A negative acute EBV-like RNA score in late blood or CSF should not be used to argue that EBV was irrelevant upstream. Conversely, a positive latency/LCL-like host program in CSF cannot by itself prove that EBV-infected or clonally descended cells produced that signal; transformation-associated programs can overlap with generic activation and tissue adaptation.

This distinction matters as EBV-directed prevention and treatment strategies develop \cite{Munz2025,Wahbeh2025}. Anti-CD20 therapy establishes the clinical importance of B cells in MS \cite{Hauser2017}, but our data do not show that its benefit is mediated by removal of an F12-like state. Trials seeking an EBV-related pharmacodynamic marker should therefore pair host transcriptomics with direct virologic measures when feasible, EBV-antigen-specific B/T cells, BCR/TCR clonotypes, chromatin accessibility, and repeated sampling before and after intervention. A dynamic panel that distinguishes early/proliferative from latency/transformation-associated host programs may be more informative than one locked score, but C10 should be viewed as hypothesis-generating for that clinical application rather than a validated assay.

The prospective negative result remains decisive for immediate biomarker claims. F12 and all external EBV programs failed the frozen Gate-3 criterion in the small early-MS cohort. The present data therefore do not support an EBV transcriptional prognostic assay. Future studies should separate \emph{etiologic attribution}, \emph{pharmacodynamic monitoring}, and \emph{prediction}; the optimal marker for one purpose need not be optimal for another.

\subsection{Limitations}
F12 was derived from only 2 experimental donors using one fixed NMF rank; rank stability across alternative decompositions was not established. The held-out-cell analysis reused the same donor libraries and F12 had the smallest held-out effect among the 3 retained factors. The original nuisance panel did not include explicit hypoxia/glycolysis programs, although C6 addressed that concern downstream. Gate 2 tested only 3 frozen factors and 2 passed. The mouse binary TBX21$^+$CXCR3$^+$ fraction is dropout-sensitive and attenuated after depth matching, so it should not be used to infer anatomical depletion; the robust anatomical result is F12 loss, not marker-fraction loss.

Human cohorts were small, particularly the 3-person LTNA group and the independent GSE138266 challenge. C8 recovered only one control at the relaxed cell threshold, so disease specificity remains unresolved. C2 odds ratios are descriptive specificity controls rather than formal tests of differences between lineages. C3 included only 5 EBV and 3 PBS evaluable mice. C7's early and LCL signatures came from the same external series and shared 22/40 source genes, so they are not independent biological mechanisms; nevertheless, only 11 shared genes remained in the C10 variance-qualified universe, and the post hoc disjoint-gene sensitivity preserved both directions after removing all source overlap. The Carter analysis used the exact 27-gene modern-symbol subset frozen in C7 rather than reconstructing the complete 38-RNA parent set after outcomes were known. C9, C10, and the disjoint-gene sensitivity, like C1--C8, are post hoc robustness/specificity analyses. Their role is to constrain and characterize the original inference, not to optimize a replacement hypothesis.

\section{Conclusion}

EBV-conditioned B-cell transcription is not represented by one durable RNA state across infection, CNS selection, and human MS compartments. The frozen F12 program is robustly depleted in human CSF and EBV-mouse brain even after coherence, metabolic, sequencing-depth, and detection controls. In human CSF, independently defined programs show calibrated stage structure: an early/proliferative program is depleted while latency-III/LCL-like programs are increased, and the early/LCL separation persists after all shared source genes are removed. This stage ordering is not identical in mouse brain, underscoring that transcriptomic state is remodeled by biological context rather than carried forward as a universal barcode.

At the same time, a statistically strong cross-system bridge showed neither B-lineage nor EBV-exposure specificity. The general lesson is therefore twofold: loss of an acute transcriptional program does not exclude upstream EBV conditioning, and transcriptomic resemblance does not establish molecular ancestry. Mechanistic and clinical studies should seek continuity in lineage, antigen specificity, virologic state, chromatin, and longitudinal response rather than relying on late-compartment RNA similarity alone.

\section*{Author Contributions}
Hasan M. Jamil: Conceptualization, Methodology, Software, Formal analysis, Investigation, Data curation,
Visualization, Writing – original draft, Writing – review and editing, Funding acquisition. Alexander Gow: Validation,
Biological interpretation, Supervision, Writing – review and editing.

\section*{Funding}
HJ's research work was supported in part by the National Institutes of Health IDeA grant P20GM103408. AG's research was funded in part by a grant from the National Multiple Sclerosis Society (RG-2305-41469).

\section*{Conflicts of Interest}
The authors declare no conflicts.

\section*{Data Availability}
All analyzed source datasets are publicly available through GEO under GSE189141, GSE299939, GSE267750, GSE133028, GSE138266, and GSE20200. The direct brain-versus-spleen analysis used the source study's deposited spleen/brain B-cell Seurat object, Zenodo DOI 10.5281/zenodo.15602185. Frozen notebooks, protocols, C1--C10 specificity/robustness outputs, the disjoint-gene sensitivity, and figure source data are included in the study reproducibility archive; final public DOI: Contact the corresponding author HJ.

\printbibliography
\end{document}